\documentstyle[psfig,aas2pp4]{article}

\slugcomment{to appear in the Astrophysical Journal} 
 
\begin{document}
 
\title{Imaging low order CO emission from the $z=4.12$ QSO  PSS 2322+1944}
   
\author{C. L. Carilli}
\affil{National Radio Astronomy Observatory, P.O. Box O, Socorro, NM,
87801, USA \\
ccarilli@nrao.edu}
\author{P. Cox }
\affil{Institut d'Astrophysique Spatial, Universit\'e de Paris XI,
91405 Orsay, France}  
\author{F. Bertoldi \& K.M. Menten}
\affil{Max-planck Institut f\"ur Radioastronomie, Auf dem H\"ugel
69, Bonn, D-53121, Germany}
\author{A. Omont}
\affil{Institut d'Astrophysique de Paris, CNRS, 98 bis boulevard
Arago, F-75014, Paris, France} 
\author{S.G. Djorgovski}
\affil{Astronomy Department, California Institute of Technology,
Pasadena, CA, 91125, USA} 
\author{A. Petric}
\affil{Astronomy Department, Columbia University, New York, NY USA}
\author{A. Beelen}
\affil{Institut d'Astrophysique Spatial, Universit\'e de Paris XI,
91405 Orsay, France} 
\author{K.G. Isaak}
\affil{Cavendish Astrophysics, University of Cambridge, 
Cambridge, UK}
\author{R.G. McMahon}
\affil{Institute of Astronomy, University of Cambridge, 
Cambridge, UK}

\begin{abstract}

We present observations of  CO 1--0 and CO 2--1 emission
from the $z = 4.12$ QSO PSS 2322+1944 using the  Very Large Array.  
The CO emission is extended on a spatial
scale of 2$''$. This extension 
could reflect the double nature of the
QSO as seen in the optical, or could be diffuse emission
with a (redshift corrected) mean brightness temperature of
2.8~K for the CO(2-1) line. We find the CO excitation conditions
are lower than in two other IR-luminous $z > 4$ QSOs,
suggesting the presence of a significant contribution from cooler, 
lower density molecular gas (n(H$_2$) $\sim 5\times 10^3$ cm$^{-3}$),
although such a conclusion is complicated by the possibility of 
differential gravitational  magnification.

\end{abstract}
 

\section {Introduction}

Two of the more important recent discoveries on
nearby galaxies are: (i)  the large majority of spheroidal galaxies
in the nearby universe contain massive black holes, and (ii) the black
hole mass correlates with the mass of the spheroid  of the parent
galaxy  (Richstone et al. 1998; Gebhardt et al. 2000;
Ferrarese \& Merritt 2000). These discoveries suggest a  
`causal connection between the formation
and evolution of the black hole and the bulge'
(Gebhardt et al. 2000), and have 
have led to the hypothesis of co-eval formation of massive
black holes and galaxy spheroids, perhaps occurring in merging galaxies
at high redshift (Franceschini et al. 1999, Blain et al. 1999,
Kauffmann \& Haenelt 
2000; Page et al. 2002).  We have undertaken an extensive
observational program 
from radio through (sub)mm wavelengths of high redshift QSOs in
order to address the interesting possibility of co-eval black hole and
spheroidal galaxy formation (Carilli et
al. 2001a, b; Omont et al. 2001, Omont et al. 2002,
Carilli, Menten, \& Yun 1999; Carilli et al. 2002). 

The $z = 4.12$ QSO PSS 2322+1944 has an $M_B = -28.1$, and was
discovered in the Digitized Palomar Sky Survey high redshift QSO
search (Djorgovski et al. in
prep.\footnote{http://www.astro.caltech.edu/~george/z4.qsos}).
Djorgovski et al. (2002) found that 2322+1944 is a double source in
the optical, with two similar spectrum components separated by about
1.5$''$. This morphology may indicate a QSO pair at $z = 4.12$, or
gravitational lensing by an intervening galaxy.

PSS 2322+1944 is the brightest (sub)mm source 
from the recent surveys of Omont et al. (2002) and
Isaak  et al. (2002), with a flux density of
$9.6\pm 0.5$ mJy at 250 GHz and $22.5\pm 2.5$ mJy at 350 GHz.
The rest frame IR spectral energy distribution (SED) 
of 2322+1944 is consistent
with thermal emission from dust at a temperature of about 47~K 
(Cox et al. 2002). The  apparent FIR luminosity 
is then $3 \times 10^{13}$ L$_\odot$.
Radio continuum emission from
2322+1944 was detected at 1.4 GHz with the VLA, with
a total flux density of 102$\pm$20 $\mu$Jy, and 
an intrinsic size of about 2$''$ (Carilli et al. 2001b). 

Cox et al. (2002) show that the radio through IR SED
for 2322+1944 is similar to that of the 
dwarf nuclear starburst galaxy M82, supporting the idea
of active star formation  co-eval with the
AGN activity of the QSO.  
The implied star formation rate, up to 
10$^3$ M$_\odot$ year$^{-1}$ if the source is
not strongly gravitationally lensed, is adequate to form
the majority of the stars in a normal spheroidal galaxy
in a relatively short time ($\le 10^8$ years). 
On the other hand, 
the global SED for 2322+1944 from cm-to-optical wavelengths is not out of
the range defined by lower redshift, lower luminosity QSOs (Carilli et
al. 2001b; Sanders et al. 1989), and hence the case for dust-heating
by star formation based on the SED alone is by no means secure.

Omont et al. (2001) point out that a 
pre-requisite for star formation at the rate considered for PSS 2322+1944
is a massive reservoir of molecular gas ($\sim 10^{11}$ M$_\odot$). 
Such reservoirs have been detected via their CO line emission
in other IR-luminous QSOs at $z > 4$ (Omont et al. 1996a;
Ohta et al. 1996; Guilloteau et al. 1997, 1999; Carilli et al. 2000; 2002).
Recently, Cox et al. (2002) have detected  emission in the 
high order CO transitions from 2322+1944. They find that
2322+1944 has the  highest apparent CO(5-4) line 
luminosity of any $z > 4$ QSO. 

In this paper we present observations of CO 1--0 and CO 2--1
emission from the PSS 2322+1944 using the  Very Large Array. 
From these data we derive 
the spatial distribution and excitation conditions of the
molecular gas.
We assume $H_o$ = 65 km s$^{-1}$ Mpc$^{-1}$, $\Omega_M = 0.3$ and 
$\Omega_\Lambda = 0.7$. 

\section{Observations}

PSS 2322+1944 was observed with the VLA in the fall of 2001
using the D configuration (1 km maximum baseline). Observations
were made of the redshifted CO 1--0 line at 22.515 GHz,
and of the CO 2--1 line at 45.035 GHz. A total of 14 hours 
was spent on each transition. 

The CO 1--0 observations at 22.5 GHz employed the
spectral line correlator mode with  two polarizations 
and 7 spectral channels of 6.25 MHz width (= 83 km s$^{-1}$). 
Due to limitations with the VLA bandwidth and
correlator, the CO 2--1 observations
at 45 GHz were done using the continuum correlator 
mode with two IFs of 50 MHz bandwidth  and two polarizations each.  
One of the IFs was centered on the CO 2--1 emission line,
while the second was centered 350 km s$^{-1}$ off the line 
in order to determine  the continuum level, if any.
This mode maximizes sensitivity to the line,
but sacrifices a measurement of the line velocity profile.
Our CO 2--1 analysis will necessarily assume the line profile
as given by the 1--0 line.

Standard amplitude and phase calibration were applied, correcting for
atmospheric opacity at high frequency, and the absolute flux density
scale was set by observing 3C~286. In the analysis below
we include a 10$\%$ systematic error due to calibration uncertainties
at high frequency.  Fast switching phase calibration
was employed (Carilli \& Holdaway 1999) with a cycle time of 3 minutes,
although subsequent inspection of the phase stability showed that
a cycle time of 15 minutes, or even longer, would have been
easily adequate to maintain phase coherence. The observations took
place at night under excellent weather conditions, with RMS phase
variations after calibration  typically less than $10^\circ$. 
The phase coherence  was checked by imaging a calibrator with 
a similar  calibration cycle time as that  
used for the target source. At all times the coherence was found to
be better than 90$\%$.

\section{Results and Analysis}

Figure 1a shows the CO 1--0 emission spectrum
from 2322+1944. A Gaussian profile was fit
to the spectrum with a resulting full width at 
half maximum (FWHM) of $200 \pm 70$ km s$^{-1}$, 
a peak flux density of $0.89\pm 0.22$ mJy, and a velocity-integrated
flux density of $0.19\pm 0.08$ Jy km s$^{-1}$.
Zero velocity in Figure 1a is defined as the center
of channel 4, corresponding to a heliocentric redshift of 
4.11956 for CO 1--0. The Gaussian fit has a central velocity of
$-22 \pm 24$ km s$^{-1}$ relative to this 
redshift, corresponding to a redshift of 
$4.1192\pm 0.0004$. For comparison, the Gaussian fit to the
CO(5-4) line has a FWHM = $273 \pm 50$ km s$^{-1}$,
and a central redshift of $4.1199 \pm 0.0008$ (Cox et al. 2002). 
 
The image of the CO 1--0 emission averaged
over channels 3, 4, and 5 is shown 
in Figure 1b,  and the average
of the continuum channels (1 and 6) is
shown in Figure 1c. The source is spatially unresolved, with
an upper limit to the (deconvolved) source size of 
3$''$ (derived from Gaussian fitting), and a peak
flux density of $0.65 \pm 0.07$ mJy (note that this peak
corresponds to the average of channels 3, 4, and 5). 
The upper limit (2$\sigma$) to continuum emission at 22 GHz is 
0.14 mJy. 

Figure 2a shows the integrated CO 2--1 emission from 2322+1944, while
2b shows the continuum channel. No continuum emission is detected to 
a 2$\sigma$ limit of 0.15 mJy. 
The CO emission is resolved spatially. A
Gaussian fit results in a deconvolved source size of FWHM = $2.1''
\times 1.6''$ with a major axis position angle of 152$^o$, and a total
flux density of $2.7\pm0.24$ mJy. 
The implied (redshift corrected) mean
brightness temperature is 2.8 K.
The velocity-integrated
flux density is $0.92 \pm 30$ Jy km s$^{-1}$, assuming a line width
as given by the CO 1--0 transition. 
The CO 2--1 image also shows emission extending about 2$''$ to the
southwest of the peak position.  This structure repeats (to
within the noise) on the three different observing days. The average
(redshift corrected) brightness temperature of this emission is $0.5
\pm 0.1$ K. We note that the observed morphology of the CO 2--1 emission 
agrees within the errors with that observed for the radio 
continuum emission at 1.4 GHz (Carilli et al. 2001b). 

The data for the velocity-integrated flux densities for various CO
transitions for 2322+1944 are plotted in Figure 3. The higher order
transition data is taken from Cox et al. (2002).  Also plotted are the
corresponding numbers for two other IR-luminous QSOs at $z > 4$,
BRI 1202--0725 at $z= 4.7$ and BRI 1335--0417 at $z= 4.4$,
for which multiple CO transitions have been observed (see Carilli et
al. 2002; Cox et al. 2002).  The data have been
normalized to the velocity-integrated line flux density 
for CO(5-4). For comparison, we
have included the CO ladder observed for the best studied nuclear
starburst galaxy M82 (G\"usten et al. 1993; Mao et al. 2000), and that
for the integrated emission from the Milky Way disk inside
the solar radius (excluding the Galactic center) as seen by COBE
(Fixsen, Bennett, \& Mather 1999).

We have fit a standard one component LVG model simulating a spherical
cloud to interpret the observed line ratios (for details see Carilli
et al. 2002).  Due to the lack of constraints, there is a degeneracy
in LVG modeling between various parameters such as the density and kinetic  
temperature (G\"usten et al. 1993).  Our modeling is not
meant as an exhaustive analysis, but is merely representative of the
types of conditions that can give rise to the observed line ratios.
Given the paucity of information, we 
adopt a kinetic temperature of 47~K, corresponding to the temperature
derived from the far-IR spectrum of thermal  
emission by warm  dust (Cox et al. 2002).  We also use the 
cosmic microwave background radiation temperature at
$z = 4.12$ of 14~K. The CO
data can be reasonably fit by a model with: N(CO) = 
$2 \times 10^{19}$ cm$^{-2}$, n(H$_2$) = $5\times10^3$ cm$^{-3}$, and
grad $V$ = 1 km s$^{-1}$ pc$^{-1}$. 
Given the inhomogeneity of the
interstellar medium in the Milky Way and nearby galaxies, 
our assumption of uniform physical conditions is crude but 
appropriate for a source with four detected lines from the 
main CO species and no information on rare isotopomers. 

\section{Discussion}

The optical QSO positions are shown as crosses on Figure 2a.  The
deconvolved size and position angle of the CO 2--1 emission from PSS
2322+1944 are consistent with the separation and position angle of the
double optical QSO components found by Djorgovski et al. (2002), as is
the observed radio continuum morphology at 1.4 GHz (Carilli et al.
2001). However, the resolution of the current VLA image is
insufficient to 
determine if the CO emission is from two small components associated
with the optical QSOs, or from more diffuse emission with a mean
(redshift corrected) brightness temperature of 2.8 K.  The more
extended emission to the southwest appears to be robust, although,
given the relatively low surface brightness, we feel this extension
needs to be confirmed with further observations.  

As discussed in Carilli et al. (2002),
the CO excitation conditions for two other IR-luminous QSOs at $z >
4$, BRI 1202--0725 and  BRI 1335--0417,
are similar to those seen in the dwarf 
nuclear starburst galaxy M82, and are consistent with a fairly high
kinetic temperature of 70 K 
and a high (mean) molecular gas density of  $2 \times 10^4$
cm$^{-3}$. For 2322+1944 we find excitation 
conditions intermediate between those seen in M82 and
those seen for the disk of the Milky Way. The data are consistent
with a kinetic temperature equal to the temperature derived from 
the far-IR SED of the thermal emission by warm dust (47 K), and 
a density for the molecular gas of $\rm n(H_2) = 5 \times 10^3$
cm$^{-3}$, ie. a factor 4 lower than that found for 1202--0725
and 1335-0417.  These results may indicate that the CO emission from
2322+1944   is dominated by gas that is more diffuse and cooler than
is seen in the other two QSOs. However, if the CO emission
from 2322+1944 is strongly lensed, then differential magnification
due to (possibly) different source-plane  spatial distributions for 
the low and high excitation molecular gas makes a detailed
excitation analysis problematic until higher resolution 
observations, and a proper lens model, are available. 

The apparent CO 1--0 luminosity, $L'$, for PSS 2322+1944
is $1.5 \times 10^{11}$ K km s$^{-1}$ pc$^2$.
The H$_2$ gas mass can be calculated from the CO luminosity
assuming a value of   X = the H$_2$ mass-to-CO(1-0)
luminosity conversion factor in M$_\odot$ (K km s$^{-1}$ pc$^2$)$^{-1}$
(Solomon et al. 1997).  We adopt a value of 
$\rm  X \simeq 0.8$, as found for Ultra-Luminous Infrared Galaxies
($L_{FIR} \sim 10^{12}$ L$_\odot$; ULIRGs) 
by Downes and Solomon (1998), leading to a 
molecular gas mass of  $1.2 \times10^{11}$ M$_\odot$.
Of course, X depends on a number of factors,
including  the excitation conditions,
and in the case of 2322+1944 the possibility of gravitational
lensing complicates the situation.

Lastly, we consider the continuum-to-line ratio: $[L_{\rm
FIR}]/[L'_{\rm CO(1-0)}]$, in units of L$_\odot$ (K km s$^{-1}$
pc$^{2}$)$^{-1}$.  Solomon et al. (1997) show that this ratio
increases with increasing $L_{\rm FIR}$, with values between 5 and 50
for Galactic Giant Molecular Clouds and nearby galaxies with $L_{\rm
FIR} \le 10^{10}$ L$_\odot$, and values between 80 and 250 for ULIRGs.
Evans et al. (2001) have extended this relationship to include
IR-excess, optically selected QSOs at $z \le 0.17$.  These QSOs have
L$_{\rm FIR}$ values ranging  from a few $\times 10^{11}$ L$_\odot$  
to  10$^{12}$ L$_\odot$.  They
find that the low $z$ QSOs typically show continuum-to-line ratios (as
defined above) between 100 and 260, with the majority of sources
toward the low end of this range.  [Note that we have converted from
their L$_{\rm IR}$ to L$_{\rm FIR}$ using a ratio of 0.52, as found by
Dale et al. (2000) for a large sample IR-selected galaxies.]
Evans et  al. (2001) also find that the QSOs with high
continuum-to-line ratios reside in optically disturbed galaxies,
ie. on-going mergers of gas rich spiral galaxies, while the sources
with low values reside in more normal-looking galaxies.

For 2322+1944 we find: $[L_{\rm FIR}]/[L'_{\rm CO(1-0)}] = 200$
L$_\odot$ (K km s$^{-1}$ pc$^{2}$)$^{-1}$.  For comparison, in BRI
1335--0417 and BRI 1202--0725 values of about 340 were found by
Carilli et al. (2002).  Carilli et al.  argued that the large
continuum-to-line ratio in these latter two QSOs, both of which have
$L_{FIR} \ge 10^{13}$ L$_\odot$, continues the trend of increasing
continuum-to-line ratio with increasing IR luminosity.  The fact that
2322+1944 has a continuum-to-line ratio more consistent with ULIRGs,
and with low redshift, lower IR luminosity QSOs, could be used to
argue for strong (factor $\sim 10$) gravitational magnification of the
emission regions, thereby lowering its IR luminosity to $\sim 10^{12}$ 
L$_\odot$. However, given the scatter in the relationship between IR
luminosity and the continuum-to-line ratio, this is at best a weak
argument in favor of gravitational lensing.

Detecting the massive reservoir of molecular gas in PSS 2322+1944
is one of the required (although not sufficient) 
elements for demonstrating
the existence of active star formation  co-eval with the
AGN activity in this high-$z$ QSO.
However, the possibility of gravitational lensing in PSS 2322+1944 
complicates the physical analysis.  An argument in favor of lensing
is the relatively low velocity dispersion for the CO emission. The
observed value of 200 km s$^{-1}$ is typical for a galaxy potential,
but is low for a group potential, as might be expected if the observed
CO line was the sum of emission from the host galaxies of a QSO pair
at $z = 4.12$. If the system is strongly gravitationally lensed, then
judicious use of a lens model could elucidate structure in the source
plane  on finer physical scales ($\le 100$ pc) than can be obtained
normally (Yun et al. 1997; Lewis et al. 2002). Sensitive, high
resolution observations of the CO 2--1 emission from 2322+1944 are
planned to confirm, and better resolve, the molecular line emitting
regions, while observations of higher order transitions at
high spatial resolution are planned to better constrain
the CO excitation conditions. 

\vskip 0.2truein 

The National Radio Astronomy Observatory (NRAO) is operated
by Associated Universities, Inc. under a cooperative agreement with the
National Science Foundation. SDG acknowledges partial support
from the Bressler Foundation. We thank A. Mahabel for assistance
with the optical astrometry, and the referee, A. Evans, for 
very helpful comments.


\clearpage
\newpage

\centerline{\bf Figure Captions}

F{\scriptsize IG}. {\bf 1a}.--- The CO 1--0 spectrum of PSS 2322+1944
as measured  at 22.515 GHz at a spatial resolution of 
3.8$''$. Zero velocity is defined as the center
of channel 4, corresponding to a heliocentric redshift of 
4.11956 for CO 1--0. Each channel is 83 km s$^{-1}$ wide, 
and the rms noise in each channel is
0.12 mJy beam$^{-1}$. The dashed line is a Gaussian
fit to the velocity profile (see section 3). \\
{\bf 1b}. The contour image of the average of channels 3, 4, and 5
from the spectrum shown in Figure 1a. The contour levels
are: -0.24, -0.12, 0.12, 0.24, 0.36, 0.48, 0.60 mJy beam$^{-1}$. 
The Gaussian restoring CLEAN beam has FWHM = 
$3.9'' \times   3.7''$ with a major axis position angle of
-46$^\circ$, as shown in the inset.
The rms noise on this image is 74$\mu$Jy beam$^{-1}$. 
The crosses in this and subsequent images show the position
of the two optical QSOs found by Djorgovski et al. (2002) \\
{\bf 1c}. The contour image of the off-line channels (1 and 6)
from the spectrum shown in Figure 1a.
The contour levels and beam are the same as Figure 1b.

F{\scriptsize IG}. {\bf 2a}.---The contour image of the average of 
the on-line IF for the CO 2--1 observations of PSS 2322+1944.
The IF bandwidth is 50 MHz (= 350 km s$^{-1}$), and is centered at
45.035 GHz, corresponding 
to a heliocentric redshift of 4.1191 GHz for CO 2--1. 
The contour levels are -0.24, -0.12, 0.12, 0.24, 0.36, 0.48, 0.60 mJy 
beam$^{-1}$. The  Gaussian restoring CLEAN beam has FWHM = 
$1.8'' \times   1.4''$ with a major axis position angle of
-28$^\circ$, as shown in the inset.
The rms noise on this image is 77$\mu$Jy beam$^{-1}$.  \\
{\bf 2b}. The contour image of the off-line IF for the 44.985 GHz 
VLA observations of 2322+1944.
The contour levels and beam are the same as Figure 2a.

F{\scriptsize IG}. {\bf 3}. --- The velocity-integrated flux densities
for CO emission from  PSS 2322+1944, as derived from
data from this paper and from Cox et al. (2002), 
are show as large filled circles with error bars.
Also plotted are the corresponding values
for the QSOs BRI 1202--0725 at $z = 4.7$
(open squares), and BRI 1335--0417 at $z = 4.4$ (open
triangles) (from Carilli et al. 2002).  The solid triangles are
the data for the CO ladder for the integrated emission from 
the Milky Way disk inside the solar radius (excluding the Galactic
center) as seen by COBE (Fixsen, Bennett, \& Mather 1999).
The open circles are the results for the
starburst nucleus of M82 (G\"usten et al. 1993; Mao et al. 2000).
The ordinate is the velocity-integrated line
flux density. The CO(5-4) line has been detected in 
all the sources, and the velocity-integrated line
flux densities for the other transitions in each source have all been 
normalized by the corresponding CO(5-4) values,  except for the Milky
Way, for which the values are normalized at CO(3-2). The
short dash line shows an LVG model with T$_{\rm kin}$ = 47 K and 
n(H$_2$) = $5 \times 10^3$ cm$^{-3}$.

\clearpage
\newpage
\begin{figure}
\vskip -0.4in
\hskip 1in
\psfig{figure=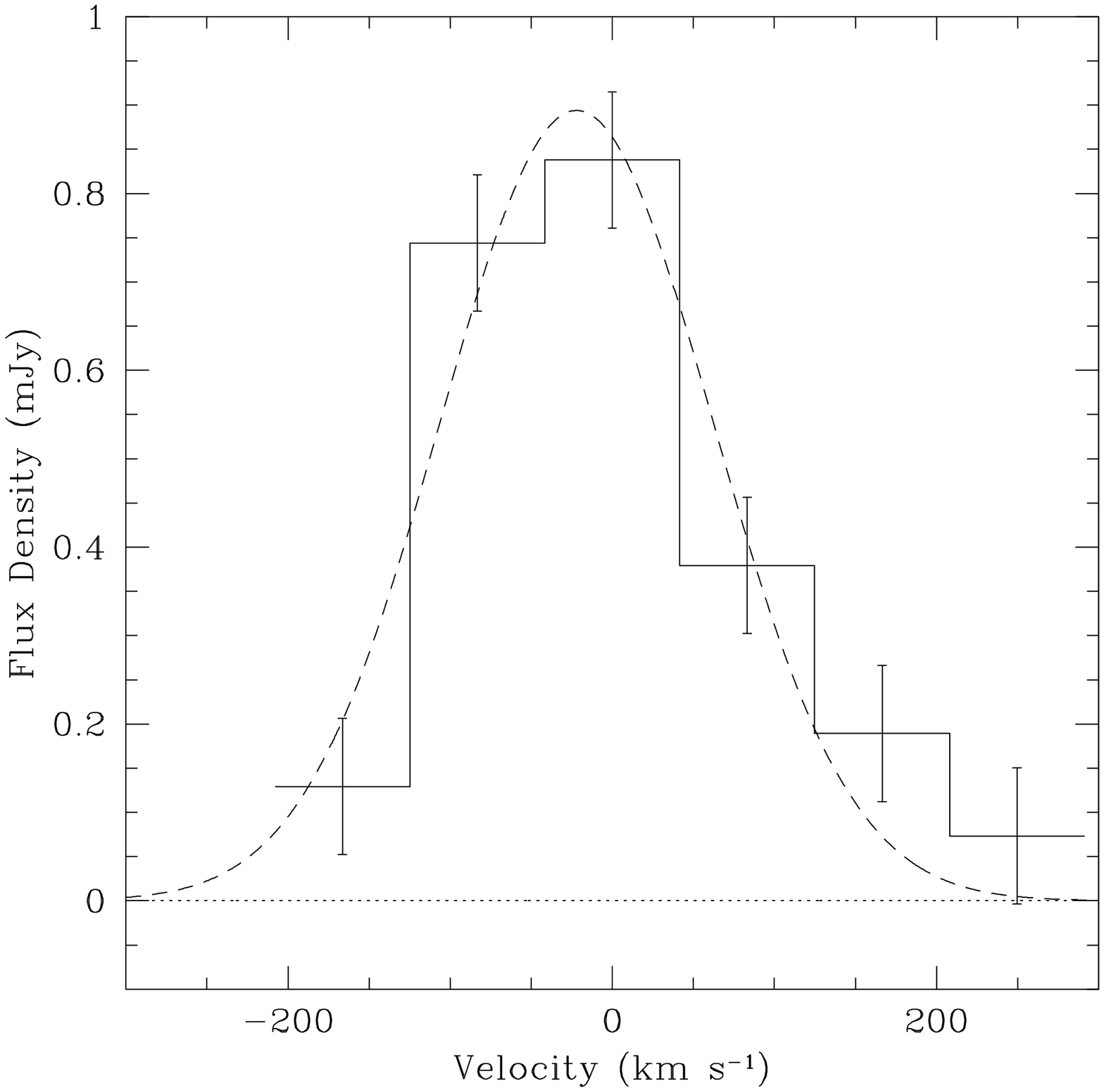,width=4in}
\psfig{figure=carilli.fig1b.ps,width=3.5in}
\vskip -3.72in
\hspace*{3.6in}
\psfig{figure=carilli.fig1c.ps,width=3.5in}
\caption{1a -- upper; 1b -- lower left; 1c -- lower right}
\end{figure}

\clearpage
\newpage

\begin{figure}
\psfig{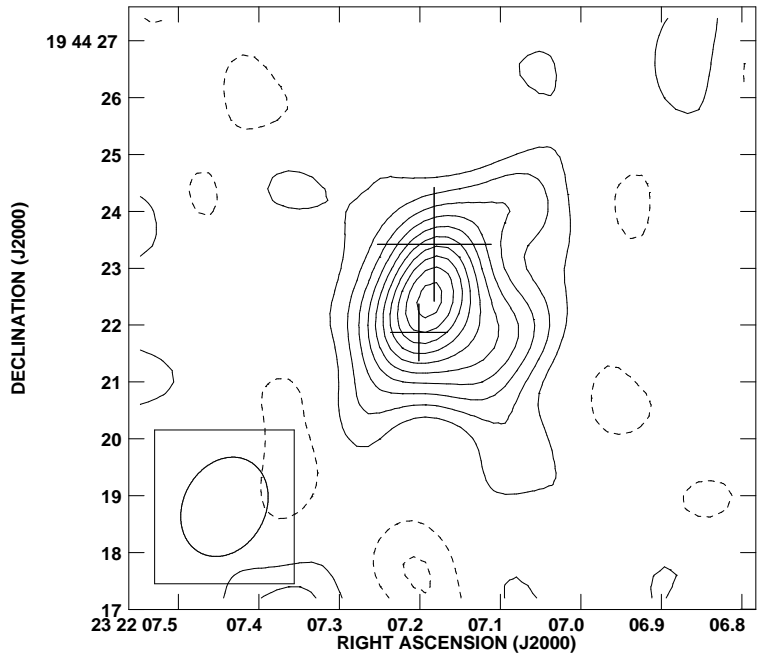}
\psfig{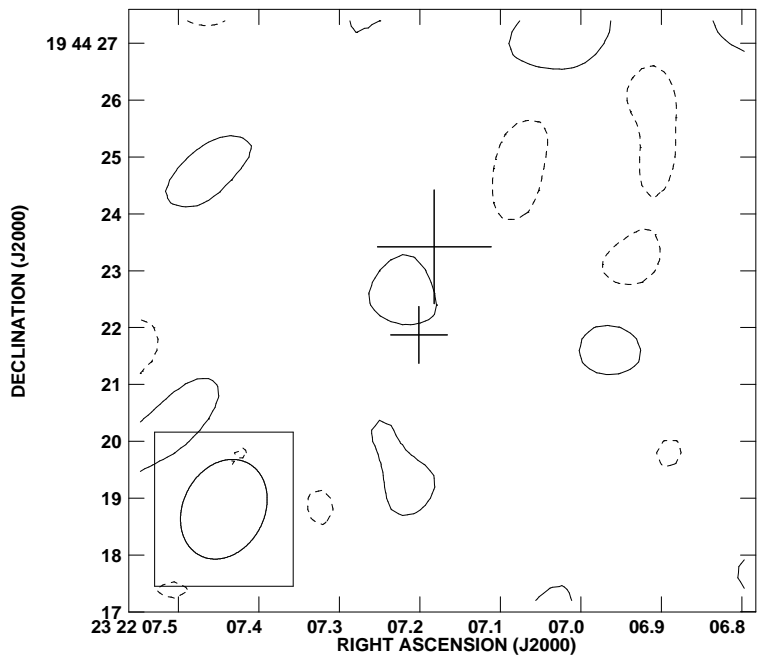}
\caption{2a -- upper; 2b -- lower}
\end{figure}

\clearpage
\newpage

\begin{figure}
\psfig{figure=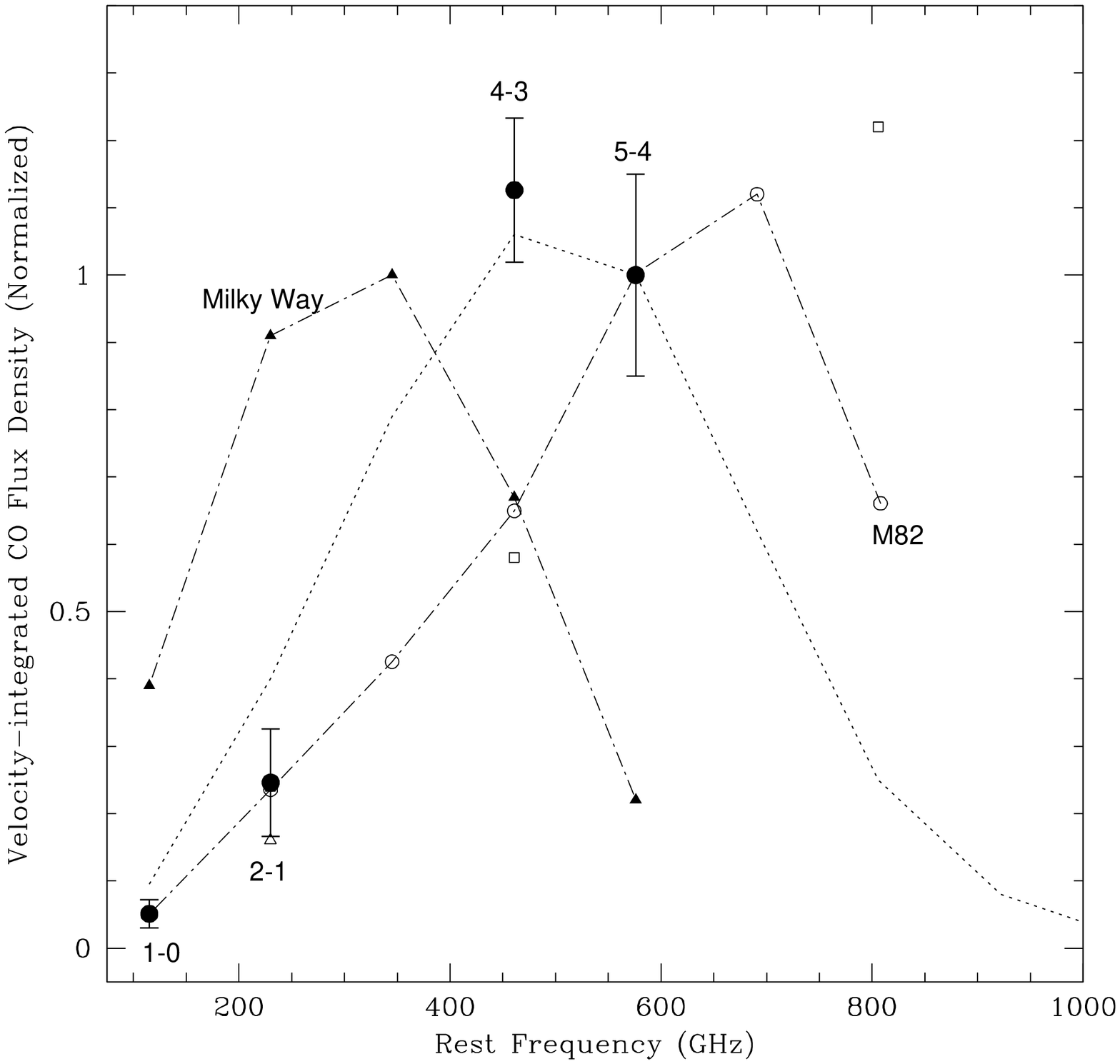,width=6in}
\caption{}
\end{figure}

\clearpage
\newpage

\end{document}